\documentclass[aps,pra,twocolumn,amsmath,amssymb,showpacs,floatfix]{revtex4-1}

\usepackage{times}

\usepackage[utf8x]{inputenc}
\usepackage[english]{babel}
\usepackage[T1]{fontenc}
\usepackage{lmodern}
\usepackage{amssymb}
\usepackage{bbold}
\usepackage{amsmath}
\usepackage{amsthm}
\usepackage[pdftex]{graphicx}
\usepackage{epstopdf}
\usepackage{float}
\usepackage{dcolumn}
\usepackage{bm}
\usepackage{color}
\usepackage{relsize,dsfont,mathrsfs,empheq,verbatim,upgreek}
\usepackage{etoolbox}
\usepackage[caption = false]{subfig}
\usepackage{braket}

\usepackage{pgfplots}

\DeclareMathOperator{\Tr}{Tr} 

\begin{document}

\title{Non-Markovianity of Quantum Brownian Motion}

\author{Simon Einsiedler}
\affiliation{Physikalisches Institut, Universit\"at Freiburg, 
Hermann-Herder-Stra{\ss}e 3, D-79104 Freiburg, Germany}

\author{Andreas Ketterer}
\affiliation{Physikalisches Institut, Universit\"at Freiburg, 
Hermann-Herder-Stra{\ss}e 3, D-79104 Freiburg, Germany}

\author{Heinz-Peter Breuer}
\affiliation{Physikalisches Institut, Universit\"at Freiburg, 
Hermann-Herder-Stra{\ss}e 3, D-79104 Freiburg, Germany}

\begin{abstract}
We study quantum non-Markovian dynamics of the Caldeira-Leggett model, a prototypical model for quantum
Brownian motion describing a harmonic oscillator linearly coupled to a reservoir of harmonic oscillators. 
Employing the exact analytical solution of this model one can determine the size of memory effects for arbitrary 
couplings, temperatures and frequency cutoffs. Here, quantum non-Markovianity is defined in terms 
of the flow of information between the open system and its environment, which is quantified through 
the Bures metric as distance measure for quantum states. This approach allows us to discuss
quantum memory effects in the whole range from weak to strong dissipation for arbitrary Gaussian initial states. 
A comparison of our results with the corresponding results for the spin-boson problem 
show a remarkable similarity in the structure of non-Markovian behavior of the two paradigmatic models.
\end{abstract}
 

\date{\today}

\maketitle

\section{Introduction}

How do memory effects manifest themselves in the dynamics of open quantum systems \cite{Breuer2002}, 
what are the characteristic features of these effects and how can they be rigorously defined 
and quantified? These questions play an important role in many applications of quantum mechanics and 
have received a tremendous amount of interest in recent years 
(see, e.g., the reviews \cite{Rivas2014a,Breuer2016a,deVega2017}). In fact, the problems of the mathematical 
definition and of the physical implications of quantum memory effects in open systems has initiated 
a large variety of interesting applications to diverse quantum systems and phenomena. Examples include 
Ising and Heisenberg spin chains and Bose-Einstein condensates \cite{Apollaro2011a,Haikka2011a,Haikka2012a}, 
quantum phase transitions \cite{Gessner2014b}, optomechanical systems 
\cite{Groblacher2015}, chaotic systems \cite{Znidaric2011a,Garcia-Mata2012a}, energy transfer 
processes in photosynthetic complexes \cite{Rebentrost2011a}, spectral line shapes and two-dimensional 
spectra \cite{Green2019}, continuous variable quantum key distribution \cite{Vasile2011b}, and quantum metrology 
\cite{Chin2012a}. Moreover, experimental realizations of quantum non-Markovianity and its control have been reported
for both photonic and trapped ion systems \cite{Liu2011a,Li2011a,Smirne2011a,Gessner2014a,Cialdi2014a,Tang2015}.

As shown in Ref.~\cite{Breuer2009b} a systematic approach to the theoretical description of memory effects 
of open systems can be based on concepts of quantum information theory. Within 
this approach memory effects are connected to the exchange of information between the open 
system and its environment. This means that Markovian, i.e. memoryless behaviour is characterized 
by a continuous loss of information, that is by a continuous flow of information from the open 
system to its surroundings, while non-Markovian dynamics is distinguished by a back flow of 
information from the environment to the open system. Employing the trace distance between 
quantum states as a measure for their distinguishably \cite{Helstrom1976,Hayashi2006} and, hence, as a 
measure for the amount of information inside the open system, Markovian quantum dynamics is
characterized by a monotonic decrease of the trace distance, while non-Markovian dynamics 
features a non-monotonic behaviour of the trace distance for a pair of quantum states.

Here, we apply these concepts to the Caldeira-Leggett model of quantum Brownian motion 
\cite{Caldeira1983,Grabert1988}. To this end, we study the integrable case of a 
harmonic oscillator, representing the open system, which is coupled to a harmonic oscillator 
reservoir, and perform a systematic analysis of the quantum non-Markovianity based on the
concept of the information flow between open system and reservoir.
However, for technical simplicity we follow the approach of Ref.~\cite{Vasile2011a} and use,
instead of the trace distance, the Bures metric \cite{Hayashi2006} as distance measure for quantum states. 
This metric is defined by means of the fidelity of quantum states, it is closely related to the trace distance
and has similar mathematical properties. In particular, trace preserving completely positive maps
are contractions for the Bures metric, a property which is important for the definition of quantum
non-Markovianity \cite{Laine2010a}. The advantage of this approach is that the Bures distance 
between two Gaussian states can easily be determined by means of an analytical expression, which
enables an efficient calculation of the non-Markovianity in a wide range of system-reservoir coupling, 
reservoir temperatures and frequency cutoffs.

The paper is organized as follows. In Sec.~\ref{sec:infoflow} we briefly recall the concepts of
information flow and non-Markovianity of the dynamics of open quantum systems. In particular, we discuss
the treatment of arbitrary Gaussian initial states as well as the application of the general approach
to the Caldeira-Leggett model. The simulation results are discussed in detail in Sec.~\ref{sec:results}.
In particular, we investigate the strong damping limit and compare our results with previous results 
obtained for the spin-boson model. Moreover, we examine the limiting case described by the
Caldeira-Leggett master equation, and study also the case of an external driving of the system or the
reservoir. Finally, in Sec.~\ref{sec:conclu} we summarise our results and draw our conclusions.

\section{Quantum information flow and memory effects}\label{sec:infoflow}

\subsection{Non-Markovianity and Gaussian initial states}

As mentioned in the Introduction we will use here the approach to quantum non-Markovianity proposed in 
Ref.~\cite{Breuer2009b} which is based on the trace distance between two quantum states $\rho_1$ and $\rho_2$
defined by
\begin{equation}
D(\rho_1,\rho_2)=\frac{1}{2}\Tr|\rho_1-\rho_2|,
\end{equation}
where $|A|=\sqrt{A^{\dagger}A}$ denotes the modulus of an operator $A$. The time evolution of the states can be 
represented by a family of trace-preserving and completely positive (CP) quantum dynamical maps $\Phi_t$,
\begin{equation}\label{family}
 \Phi = \Big\{ \Phi_t \mid \Phi_0 = {\rm id}, \, t \in [0,t_{\rm max}] \Big\},
\end{equation}
such that $\rho_{1,2}(t)=\Phi_t\rho_{1,2}(0)$ represent the states at time $t\in [0,t_{\rm max}]$. 
In view of the interpretation of the trace 
distance in terms of the distinguishability of the quantum states, a dynamical decrease of $D(\rho_1(t),\rho_2(t))$ can be 
interpreted as a loss of information from the open system into the environment characteristic of Markovian dynamics and, 
vice versa, any increase of $D(\rho_1(t),\rho_2(t))$ as a flow of information from the environment back to the system 
signifying memory effects and non-Markovian behavior. Since trace-preserving positive and completely positive maps are 
contractions for the trace distance, any P-divisible or CP-divisible dynamics \cite{Breuer2016a} provides an example of a 
monotonically decreasing trace distance and, hence, of a Markovian dynamics \cite{Chruscinski2011a,Wissmann2015a}. 
In particular, any dynamics given by a semigroup with a
generator in Lindblad form is Markovian according to this definition. 

The degree of memory effects of a given dynamics can be quantified by integrating the total information backflow
from the environment to the system leading to the non-Markoviantiy measure \cite{Breuer2009b}
\begin{equation} \label{Nonmark}
\mathcal{N}(\Phi) = \max_{\rho_{1,2}(0)} \int_{\sigma > 0} \text{d}t \ \sigma(t),
\end{equation}
where 
\begin{equation}
\label{Tracedistrate}
\sigma(t)=\frac{d}{dt}D\big(\Phi_t\rho_{1}(0),\Phi_t\rho_{2}(0)\big),
\end{equation}
and the integration is taken over all time intervals of the full interval $[0,t_{\rm max}]$ in which $\sigma(t)>0$. 

In this paper we will investigate the time evolution of Gaussian initial states, which preserve their Gaussianity under the 
action of the time evolution operators generated by the quadratic Caldeira-Leggett Hamiltonian. 
As explained in the Introduction, instead of the trace distance we will consider here the Bures distance \cite{Hayashi2006} 
of quantum states which leads to a simple expression for Gaussian states in terms of the first two moments of the position 
and momentum operators of the open system \cite{Vasile2011a}. To this end, we introduce the quantum fidelity defined by 
\begin{equation}
\mathcal{F}(\rho_1,\rho_2)=\left[\Tr\left\{\sqrt{\sqrt{\rho_1}\rho_2 \sqrt{\rho_1}}\right\}\right]^2,
\end{equation}
which is equal to one if and only if $\rho_1=\rho_2$ and equal to zero if and only if $\rho_1$ and $\rho_2$ have orthogonal 
support. We introduce the dimensionless quadrature operators $X=(\hat{a}+\hat{a}^{\dagger})/\sqrt{2}$ and 
$P= i(\hat{a}^{\dagger}-\hat{a})/\sqrt{2}$, which we can arrange in the vector operator 
$\mathbf{R}=(X,P)^T$.
The fidelity can then be expressed in terms of the first two moments of the two single mode Gaussian states 
$\rho_1$ and $\rho_2$ as
\begin{equation}
\label{Fidelitygauss}
\mathcal{F}(\rho_1,\rho_2)=\frac{2}{\sqrt{\Delta+\delta}-\sqrt{\delta}}
e^{-\frac{1}{2}(\mathbf{d}_2-\mathbf{d}_1)^T(\bm{\sigma}_1+\bm{\sigma}_2)^{-1}(\mathbf{d}_2-\mathbf{d}_1)}, 
\end{equation}
where 
\begin{align}
\Delta&=4 \det(\bm{\sigma}_1+\bm{\sigma}_2),\\
\delta&=16\left[\det(\bm{\sigma}_1)-\frac{1}{4}\right]\left[\det(\bm{\sigma}_2)-\frac{1}{4}\right].
\end{align}
Here, $\bm{\sigma}_{1,2}$ are the covariance matrices of the states with the elements
\begin{equation}
 (\bm{\sigma}_{1,2})_{ij}=\frac{1}{2}\langle \{R_i,R_j\} \rangle_{\rho_{1,2}}-\langle R_i \rangle_{\rho_{1,2}} 
 \langle R_j \rangle_{\rho_{1,2}}
\end{equation}
and 
\begin{equation}
 \mathbf{d}_{1,2} = \langle \mathbf{R}\rangle_{\rho_{1,2}}
\end{equation} 
represent the mean values of the two states. 

Now, the Bures distance is defined with help of the fidelity by means of
\begin{equation}
\label{Buresdist}
D_B(\rho_1,\rho_2)=\sqrt{2-2\sqrt{\mathcal{F}(\rho_1,\rho_2)}}.
\end{equation}
We note that, like the trace distance, the Bures distance is contracting under any trace-preserving (completely) 
positive map. Moreover, the Bures distance and the trace distance are related by the inequalities \cite{EntanglGeom}
\begin{align} \label{BuresTraceineq}
\sqrt{2-2\sqrt{1-D(\rho_1,\rho_2)^2}} \leq D_B(\rho_1,\rho_2) \leq \sqrt{2D(\rho_1,\rho_2)}.
\end{align}

The measure for non-Markovianity \eqref{Nonmark} involves a maximization over all pairs of initial states
in order to make the expression a functional of the family $\Phi$ of dynamical maps given in Eq.~\eqref{family}. 
In the following we will omit this
maximization and consider a fixed pair of initial states. The measure thus describes the degree of memory 
effects for the specific process defined by the dynamical map and a certain pair of initial conditions.
Summarizing, in the following we will investigate the quantity
\begin{equation} \label{Nm_witness}
\mathcal{N}(\Phi)=\int_{\sigma>0} \text{d}t \ \sigma(t),
\end{equation}  
where 
\begin{equation}
 \sigma(t)=\frac{d}{dt}D_B\big(\rho_1(t),\rho_2(t)\big).
\end{equation}

\subsection{Application to the Caldeira-Leggett model}\label{sec:CL-model}

The Caldeira-Leggett model describes an open quantum system linearly coupled via the position coordinate 
to an infinite bath of quantum harmonic oscillators. We assume that the open system also represents a harmonic
oscillator such that the Hamiltonian is given by \cite{Caldeira1983}
\begin{align}
\label{ham}
H=&\underbrace{\frac{1}{2m}p^2+\frac{1}{2}m\omega_0^2 x^2}_{\text{$H_S$}} 
+\underbrace{\sum_{n} \left(\frac{1}{2m_n}p_n^2+\frac{1}{2}m_n\omega_n^2 x_n^2 \right)}_{\text{$H_B$}}\nonumber\\ 
&\underbrace{-x\sum_n \kappa_n x_n}_{\text{$H_I$}}+\underbrace{x^2 \sum_{n} 
\frac{\kappa_n^2}{2m_n\omega_n^2}}_{\text{counterterm}}. 
\end{align}
The coordinate and momentum operators of the central oscillator and the bath oscillators are $x$, $p$ and $x_n$, $p_n$, 
respectively. The first two terms $H_S$ and $H_B$ are the energy contributions of the open system and the bath 
oscillators, respectively. The third part is the interaction term between the system and the bath. The coupling strength 
between the system and the $j$-th bath oscillator is represented by $\kappa_j$. Furthermore a counter term is typically 
introduced in the Hamiltonian, which accounts for a renormalization of the central oscillator frequency due to the interaction 
with the bath \cite{Grabert1988}.

The bath and interaction characteristics are fully described by the so-called spectral density
\begin{equation}
\label{specdens}
J(\omega)=\sum_{n} \frac{\kappa_n^2}{2 m_n \omega_n}\delta(\omega-\omega_n).
\end{equation}
In the limit of infinitely many bath oscillators the spectral density may be represented by a smooth function of frequency. 
In the following we choose an Ohmic spectral density with a Lorentz-Drude cutoff of the form
\begin{equation}
J(\omega)=\frac{2m\gamma}{\pi}\omega\ \frac{\Omega^2}{\Omega^2+\omega^2}.
\end{equation}
We call $\gamma$ the coupling strength and $\Omega$ the frequency cutoff.

\begin{figure}
\centering
\includegraphics[width=0.48\textwidth]{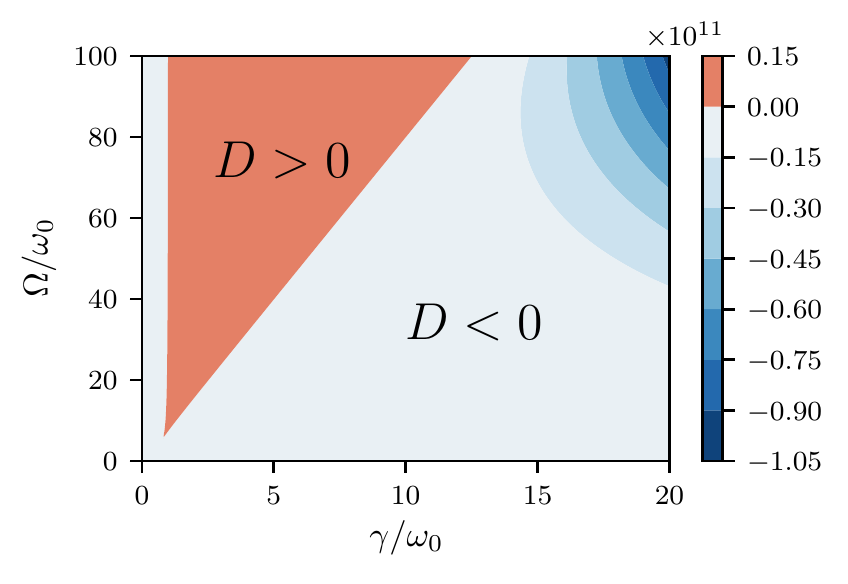}
\caption[Discriminant of the cubic equation]{The discriminant of the cubic equation \eqref{roots} as function of the
parameters $\Omega$ and $\gamma$. Blue colours indicate negative values and red colours positive values of the 
discriminant.}
\label{disc}
\end{figure}

In order to calculate the time evolution of the Bures distance for Gaussian initial states we need to find the time evolution of 
the first two moments of the momentum and position operators of the open system. Those in turn can be calculated by 
finding the time evolution of the position and momentum operators in the Heisenberg picture. From the Heisenberg 
equation of motions of $x(t)$, $x_n(t)$, $p(t)$ and $p_n(t)$ generated by the total Caldeira-Leggett Hamiltonian it is 
possible to derive the following operator valued integro-differential equation for the position operator \cite{Breuer2002}
\begin{equation}
\label{quantlangevin}
\ddot{x}(t)+\omega_0^2 x(t)+\frac{d}{dt}\int\limits_0^t dt^{\prime} \gamma(t-t^{\prime})x(t^{\prime})=\frac{1}{m}B(t),
\end{equation}
where 
\begin{equation}
B(t)=\sum_n \kappa_n \sqrt{\frac{1}{2 m_n \omega_n}}\left(e^{-i\omega_n t} a_n+e^{+i\omega_n t} a_n^{\dagger}\right)
\end{equation} is the stochastic force and 
\begin{equation}
\gamma(t)=\frac{2}{m}\int_0^{\infty}d\omega \frac{J(\omega)}{\omega}\cos(\omega t)
\end{equation} 
the damping kernel. This equation can be solved by Laplace transformation with the solution
\begin{equation} \label{generalsol}
 x(t) = G_1(t)x(0)+G_2(t)\frac{p(0)}{m}+\frac{1}{m}\int\limits_0^t dt^{\prime} G_2(t-t^{\prime})B(t^{\prime}).
\end{equation}
The solution for the time evolution of the momentum operator is determined by $p(t)=m\frac{d}{dt}x(t)$.
The functions $G_1(t)$ and $G_2(t)$ are the solutions of the homogeneous part of Eq.~\eqref{quantlangevin}
corresponding to the initial conditions $G_1(0)=1$, $\dot{G}_1(0)=0$ and
$G_2(0)=0$, $\dot{G}_2(0)=1$, respectively.  These Green's functions can be obtained by Laplace transformation
which yields
\begin{equation}
\label{laplaceintdiffeq}
\hat{G}_2(z)=\frac{\hat{G}_1(z)}{z}=\frac{1}{z^2+z\hat{\gamma}(z)+\omega_0^2}.
\end{equation}
The Laplace transform of the damping kernel corresponding to a Lorentz-Drude spectral density reads
\begin{equation}
\hat{\gamma}(z)=2\gamma \frac{\Omega}{z+\Omega}
\end{equation}
and, hence, the Laplace transform of $G_2(t)$ is given in this case by
\begin{equation}
\label{LaplaceofG_2}
\hat{G}_2(z)=\frac{z+\Omega}{z^3+\Omega z^2+(\omega_0^2+2\gamma\Omega)z+\omega_0^2\Omega}.
\end{equation}
With help of the residual theorem one performs the back transformation and arrives at
\begin{equation}
G_2(t)=\sum_{i=1}^3 d_i e^{z_it},
\end{equation}
where $z_i$ are the poles of $\hat{G}_2(z)$ determined by the roots of the cubic polynomial in the denominator in 
Eq. \eqref{LaplaceofG_2},
\begin{equation}\label{roots}
 z^3+\Omega z^2+(\omega_0^2+2\gamma\Omega)z+\omega_0^2\Omega = 0.
\end{equation}
The discriminant of this cubic equation in dependence on the coupling strength $\gamma$ and the frequency cutoff 
$\Omega$ is depicted in Figure \ref{disc}.

In the blue region the discriminant is negative and hence there exist two complex solutions, while in the red region the 
discriminant is positive which indicates the existence of only real solutions.
The factors $d_i$ are given by
\begin{equation}
d_i=\prod_{j\neq i}\frac{z_i+\Omega}{z_j-z_i}.
\end{equation}
We assume a factorizing initial condition $\rho(0)=\rho_S\otimes \rho_B$, where $\rho_S$ is an arbitrary single mode 
Gaussian state and $\rho_B=\exp\big(-H_B/(k_B T)\big)/Z_B$ is a thermal equilibrium state of the bath. From this initial 
conditions we obtain the time evolution of the mean values 
\begin{align}
\label{momentx}
\langle x(t) \rangle &=G_1(t)\langle x(0) \rangle +\frac{1}{m}G_2(t)\langle p(0) \rangle ,\\
\label{momentp}
\langle p(t) \rangle &=m\dot{G}_1(t)\langle x(0) \rangle +\dot{G}_2(t)\langle p(0)\rangle ,
\end{align}
and the elements of the covariance matrix 
\begin{align}
\label{timeevSigx}
\sigma_{xx}(t)&=G_1^2(t)\sigma_{xx}(0)+\frac{1}{m^2}G_2^2(t)\sigma_{pp}(0)\\
&\qquad+\frac{1}{m}G_1(t)G_2(t)\sigma_{xp}(0)+\langle I^2_x(t)\rangle, \nonumber \\
\label{timeevSigp}
\sigma_{pp}(t)&=m^2\dot{G}_1^2(t)\sigma_{xx}(0)+\dot{G}_2^2(t)\sigma_{pp}(0)\\
&\qquad+m\dot{G}_1(t)\dot{G}_2(t)\sigma_{xp}(0)+\langle I^2_p(t)\rangle, \nonumber\\
\label{timeevSigpx}
\sigma_{px}(t)&=m\dot{G}_1(t)G_1(t)\sigma_{xx}(0)+\frac{1}{m}\dot{G}_2(t)G_2(t)\sigma_{pp}(0) \nonumber \\
&\qquad+\big(\dot{G}_1(t)G_2(t)+\dot{G}_2(t)G_1(t)\big)\sigma_{xp}(0) \nonumber\\
&\qquad+\langle I^2_{px}(t)\rangle.
\end{align}
The crucial part in explicitly evaluating the covariance matrix for each time step is to calculate the noise contributions, 
which are of the form
\begin{widetext}
\begin{align}
\label{noise1}
\langle I_{xx}(t)\rangle&=\frac{1}{2m^2}\int\limits^t_0 ds\int\limits^t_0 ds^{\prime}G_2(t-s)G_2(t-s^{\prime})  \langle 
\{B(s),B(s^{\prime})\}\rangle,  \\
\label{noise2}
\langle I_{pp}(t)\rangle&=\frac{1}{2}\int\limits^t_0 ds\int\limits^t_0 ds^{\prime}\dot{G}_2(t-s)\dot{G}_2(t-s^{\prime})   \langle 
\{B(s),B(s^{\prime})\}\rangle,\\
\label{noise3}
\langle I_{px}(t)\rangle&=\frac{1}{2m}\int\limits^t_0 ds\int\limits^t_0 ds^{\prime}\dot{G}_2(t-s)G_2(t-s^{\prime})   \langle 
\{B(s),B(s^{\prime})\}\rangle.
\end{align}
\end{widetext}
For factorizing initial conditions the correlation function $\langle \{B(t),B(t^{\prime})\}\rangle$ of the stochastic force equals 
the so called noise kernel $D_1(t-t^{\prime})$, which has the following connection to the spectral density
\begin{equation}
\label{correlationfunction}
D_1(t-t^{\prime})=2\int\limits_0^{\infty}d\omega J(\omega) \coth\left(\frac{\omega}{2k_B T}\right)\cos\left[\omega 
(t-t^{\prime})\right].
\end{equation}
For the Lorentz Drude spectral density it can be explicitly calculated to be
\begin{equation}
\label{NoisekernelLD}
D_1(t-t^{\prime})=4m\gamma k_B T \Omega^2 \sum_{n=-\infty}^{\infty}\frac{\Omega e^{-\Omega |t-t^{\prime}|}-|\nu_n|e^{-
\nu_n |t-t^{\prime}|}}{\Omega^2-\nu_n^2},
\end{equation} 
where $\nu_n=2\pi n k_B T$ are the so-called Matsubara frequencies. 
The noise contributions can then be brought in the form
\begin{equation}
\label{infiniteNoiseseries}
\langle I_{\alpha}(t)\rangle=\frac{2\gamma k_B T \Omega^2}{m}\sum_{n=-\infty}^{\infty}\frac{1}{\Omega^2-\nu_n^2} 
\mathcal{I}_{\alpha}(t,n),
\end{equation}
where the index $\alpha$ stands for $xx$, $pp$ or $px$.
We show in Appendix \ref{appA} how to explicitly calculate the factors $ \mathcal{I}_{\alpha}(t,n)$.
With this it is possible to determine the time evolution of the elements of the covariance matrix by numrically
calculating the infinite series which appears in the noise contributions. This series turns out to converge rapidly for medium 
and high temperatures. Also for low temperatures evaluating the series numerically is preferable to a numerical integration 
of Eqs. (\ref{noise1})-(\ref{noise2}).

It is also possible to derive the time evolution of the first two moments for the Caldeira-Leggett master equation
\begin{align}
\label{QBMmastereq}
\frac{d}{dt}\rho_S(t)=-i\left[H_S,\rho_S(t)\right]-i \gamma\left[x,\left\{p,\rho_S(t)\right\}\right]\nonumber\\
-2m\gamma k_B T\left[x,\left[x,\rho_S(t)\right]\right]
\end{align}
which can be derived within the Born-Markov approximation under the conditions
$\gamma, \omega_0 \ll \text{Min}\{\Omega,2\pi k_B T\}$.
From this master equation one can derive a system of coupled differential equations for the first two moments.
It turns out that the solution of this coupled differential equations approximates the exact solution of the model
under the under the conditions $\gamma, \omega_0 \ll \Omega \ll 2\pi k_B T$. 
The mean values have the same structure as in Eqs. \eqref{momentx} and \eqref{momentp}
with the Green's functions
\begin{align}
\label{quantlangevinhomo1}
G_1(t)&=\Big[\frac{\gamma}{\nu}\sin(\nu t)+\cos(\nu t)\Big] e^{-\gamma t},\\
\label{quantlangevinhomo2}
G_2(t)&=\frac{1}{\nu}\sin(\nu t) e^{-\gamma t},
\end{align}
where $\nu=\sqrt{\omega_0^2-\gamma^2}$.
The elements of the covariance matrix have the same structure as in Eqs.~\eqref{timeevSigx}-\eqref{timeevSigpx}
with the Green's functions \eqref{quantlangevinhomo1} and \eqref{quantlangevinhomo2}, and the 
noise contributions
\begin{widetext}
\begin{align}
\langle I_{xx}(t) \rangle&\approx\frac{k_BT}{m \omega_0^2}\Bigg[1-e^{-2\gamma t}\Big (1+\frac{\gamma}{\nu}\sin(2\nu t)
+2\frac{\gamma^2}{\nu^2}\sin^2(\nu t)\Big)\Bigg], \\
\langle I_{pp}(t) \rangle &\approx m k_B T \Bigg[1-e^{-2\gamma t}\Big(\frac{\omega_0^2}{\nu^2}-\frac{\gamma^2}
{\nu^2}\cos(2\nu t)-\frac{\gamma}{\nu}\sin(2 \nu t)\Big)\Bigg], \\
\langle I_{px}(t) \rangle &\approx 2\gamma k_B T G_2^2(t).
\end{align}
\end{widetext}

\section{Simulation results and discussion}\label{sec:results}

\subsection{Time evolution of the Bures distance}\label{sec:bures}

With help of the methods developed in the previous section we are now in the position to calculate the time evolution 
of the first two moments of the position and momentum operators of the open system. As initial states we consider here
two pure coherent states,
\begin{equation}
 \rho_1(0) = |\alpha_1\rangle\langle\alpha_1|, \qquad \rho_2(0) = |\alpha_2\rangle\langle\alpha_2|,
\end{equation}
where
\begin{equation} 
 \alpha_1 = \sqrt{\frac{m\omega_0}{2}} \langle x_1 \rangle, \qquad 
 \alpha_2 = \sqrt{\frac{m\omega_0}{2}} \langle x_2 \rangle.
 \end{equation}
To explicitly evaluate the noise contributions appearing in the covariance matrix elements we need to truncate the infinite 
series in Eq. \eqref{infiniteNoiseseries} when sufficient accuracy is reached. For all results that we present in the following, 
we truncated the infinite series in the noise contribution after a number of terms $N_c$ such that the difference of the partial 
sums $\langle I_{\alpha}(t)\rangle_{(k)}$ satisfy
\begin{equation}
|\langle I_{\alpha}(t)\rangle_{(N_c)}-\langle I_{\alpha}(t)\rangle_{(N_c-1)}|<10^{-3}
\end{equation}
for every calculated time step. From the time evolution of the first two moments we are then able to compute the time 
evolution of the Bures distance with help of Eqs. \eqref{Fidelitygauss} and \eqref{Buresdist}. 

\begin{figure}
\centering
\includegraphics[]{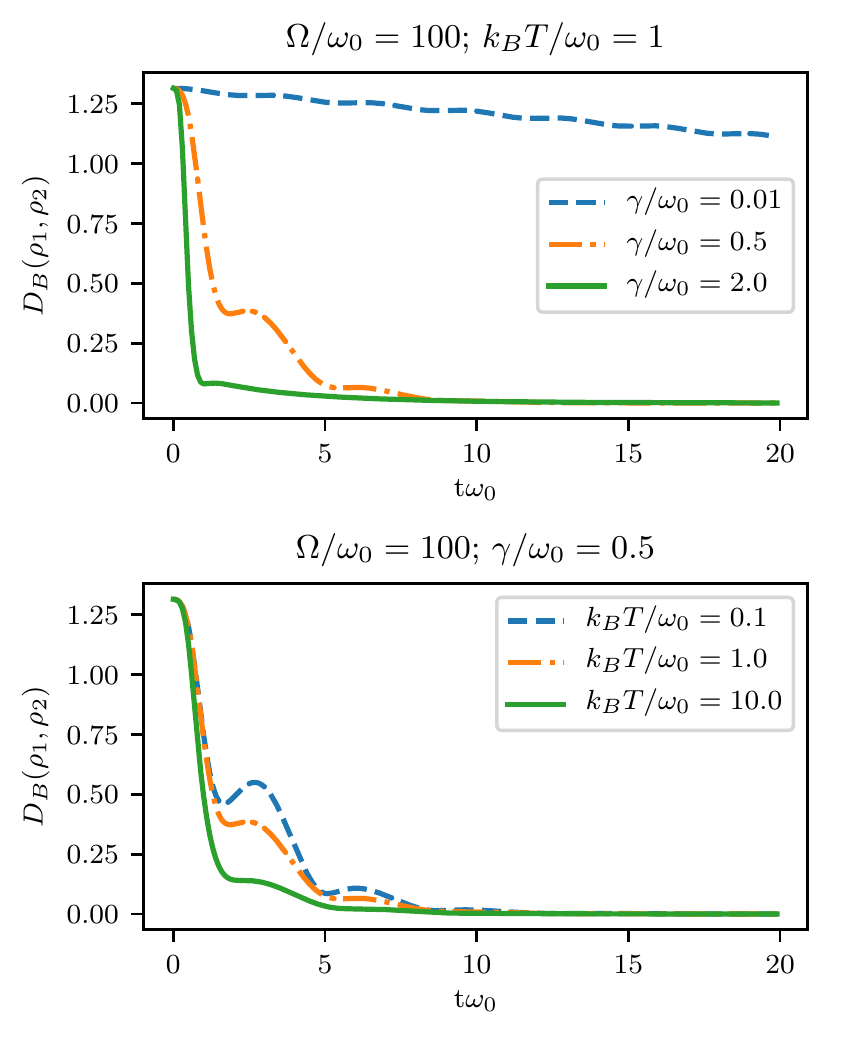}
\caption[Time evolution of the Bures distance, high cutoff (solution via Heisenberg eq.o.m.) ]{Time evolution of the Bures 
distance for different parameter combinations of the coupling strengths $\gamma$ and  initial bath temperatures $T$ for a 
cut off $\Omega/\omega_0=100$. The initial states $\rho_1(0)$ and $\rho_2(0)$ are two coherent states with  
$\sigma_{x_{1,2}x_{1,2}} m\omega_0 = \sigma_{p_{1,2}p_{1,2}} /(m\omega_0)=0.5$, $\sigma_{p_{1,2}x_{1,2}}=0$ 
and displacements $\langle x_1\rangle\sqrt{2m\omega_0}=-1$, $\langle x_2\rangle\sqrt{2m\omega_0}=1$, as well as 
$\langle p_1\rangle=\langle p_2\rangle=0$.}
\label{burestimeevom100}
\end{figure}

In Figure \ref{burestimeevom100} the time evolution of the Bures distance between two initial coherent states 
is plotted for low, medium and high coupling strengths $\gamma$ and initial bath temperatures $T$ for 
the cutoff frequency $\Omega/\omega_0=100$. The Bures distance starts in both plots nearly at its maximal value 
$\sqrt{2}$, indicating an almost maximal distinguishability between the initial states which are nearly orthogonal,
and decreases to zero during the time evolution. Furthermore we observe clear non-monotonic behaviour of the Bures 
distance. We can already see from these plots, that the bumps appearing are stronger for intermediate couplings and 
smaller for high and low couplings. Furthermore, we see that the non-monotonicity gets stronger by lowering the initial 
bath temperature.

\begin{figure}
\centering
\includegraphics[]{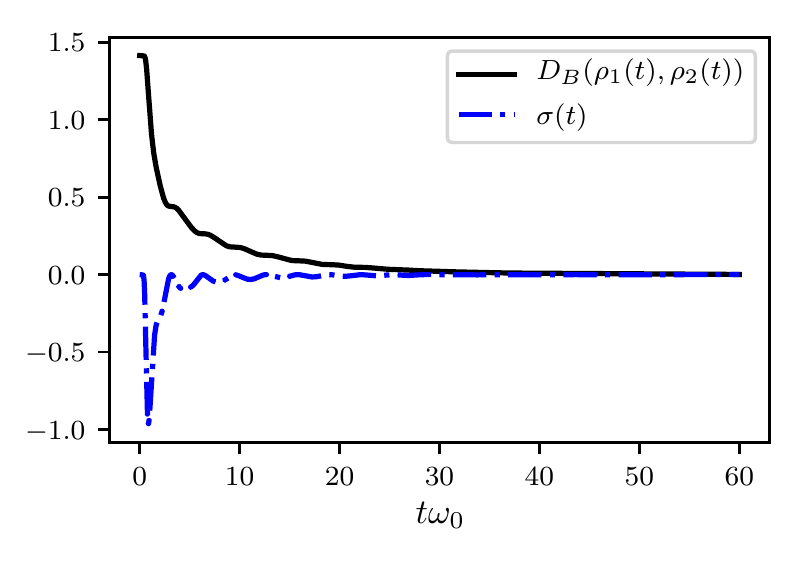}
\caption{Black line: time evolution of the Bures distance obtained via the Caldeira-Leggett master equation. 
Blue line: time derivative of the Bures distance. The coupling strength is $\gamma/\omega_0=0.1$ and the temperature 
is $k_B T/\omega_0=100$. The initial states are two coherent states with variances 
$\sigma_{x_{1,2}x_{1,2}}m\omega_0 = \sigma_{p_{1,2}p_{1,2}}/(m\omega_0)=0.5$, $\sigma_{p_{1,2}x_{1,2}}=0$ 
and displacements $\langle x_1\rangle\sqrt{2m\omega_0}=-3$, $\langle x_2\rangle\sqrt{2m\omega_0}=3$, 
as well as $\langle p_1\rangle=\langle p_2\rangle=0$.  }
\label{Clmaster_bures}
\end{figure}

We compare this to the limiting case $\gamma, \omega_0 \ll \Omega \ll k_BT$, where the dynamics of the open 
system can be described by the Caldeira-Leggett master equation as mentioned in the previous section. 
In Figure \ref{Clmaster_bures} we show the time evolution of the Bures distance obtained from the Caldeira-Leggett 
master equation and its time derivative $\sigma(t)$ for the coupling strength $\gamma/\omega_0=0.1$ and 
the initial bath temperature $k_B T/\omega_0=100$. The initial states are two coherent states with displacements 
$\langle x_1\rangle\sqrt{2m\omega_0}=-3$, $\langle x_2\rangle\sqrt{2m\omega_0}=3$, as well as 
$\langle p_1\rangle=\langle p_2\rangle=0$. We observe that $\sigma(t)$ is always negative. This implies that 
the Bures distance is monotonically decreasing and, hence, that our non-Markovianity measure of the dynamics is zero.
Note that this fact is not obvious from the beginning since the Caldeira-Leggett master equation is not in Lindblad form.

\subsection{Non-Markovianity} \label{sec:non-Markoviantiy}

\begin{figure}[h]
\centering
\includegraphics[]{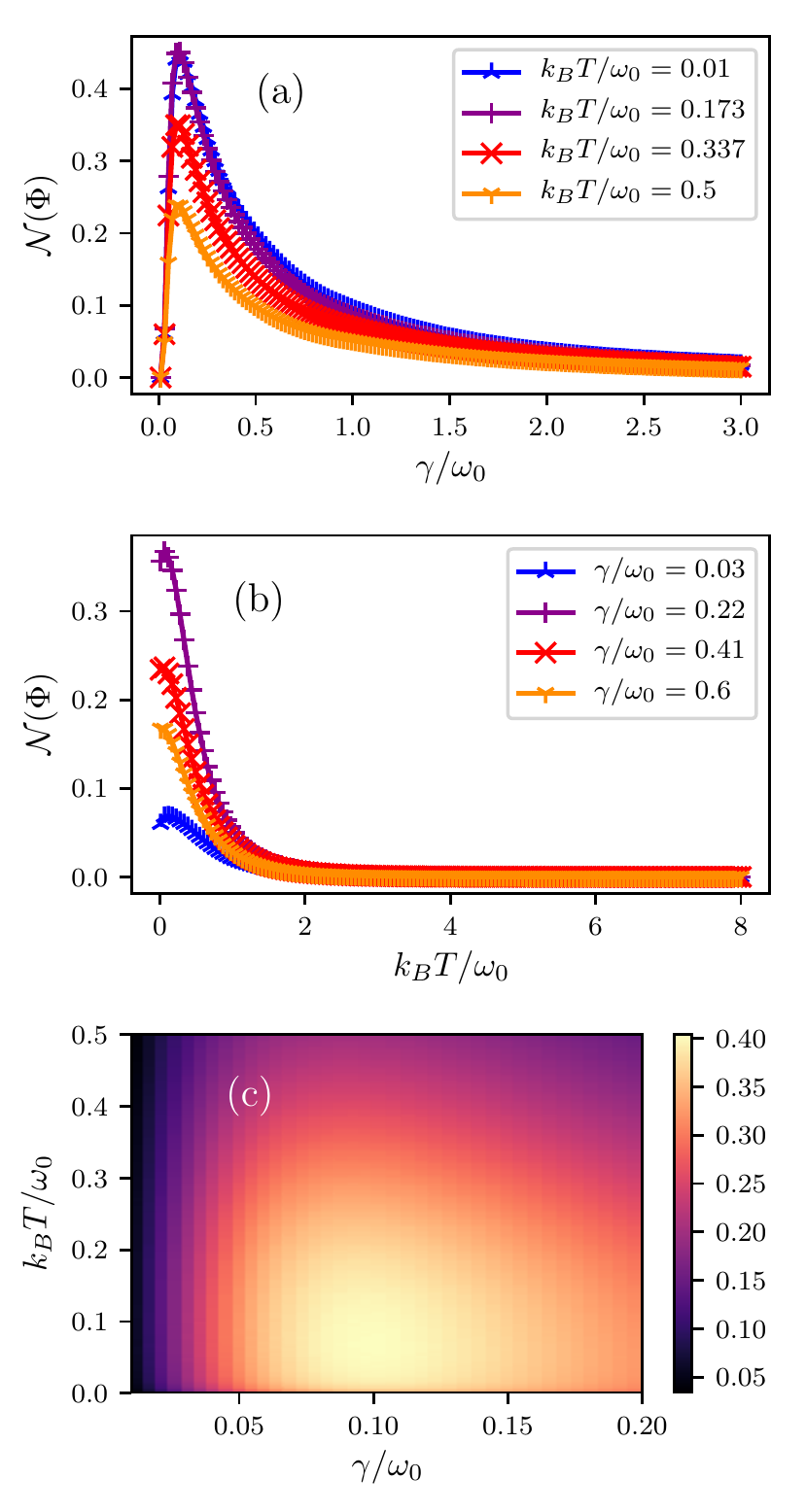}
\caption{(a)  $\mathcal{N}$ in dependence on the coupling strength $\gamma$ for different temperatures $T$. 
(b) $\mathcal{N}$ in dependence on the temperature $T$ for different coupling strengths $\gamma$. The maximal 
evaluated time is $t_{\text{max}}\omega_0=40$. The displacements are $\langle x_1\rangle\sqrt{2m\omega_0}=-3$, 
$\langle x_2\rangle\sqrt{2m\omega_0}=3$, as well as $\langle p_1\rangle=\langle p_2\rangle=0$.
(c) $\mathcal{N}$ in dependence of the coupling strength $\gamma$ and the initial bath temperature $T$. The maximal 
evaluated time is $t_{\text{max}}\omega_0=25$. The displacements are $\langle x_1\rangle\sqrt{2m\omega_0}=-1$, 
$\langle x_2\rangle\sqrt{2m\omega_0}=1$, as well as $\langle p_1\rangle=\langle p_2\rangle=0$. In all plots the cutoff 
frequency is $\Omega/\omega_0=100$.}
\label{longNMgamT_varT}
\end{figure}

We continue with our main goal of this paper to characterize the non-Markovianity measure \eqref{Nm_witness} in 
dependence on various parameters in the Caldeira-Leggett model. The numerical evaluation of the non-Markovianity 
measure is done as follows.
Given that we calculated the time evolution of the Bures distance on a grid of $N$ points of time $t_l=l\Delta t$ ($l=1,...,N$) 
with grid size $\Delta t$, we can numerically evaluate the non-Markovianity measure by
\begin{equation}
\mathcal{N}(\Phi)\approx\sum_{l=0}^{N-1}\left[ D_B(t_{l+1})-D_B(t_{l})\right],
\end{equation}
where the sum is extended over all $l$ for which $D_B(t_{l+1})-D_B(t_{l})>0$ holds.
From this it is obvious that the precision of the numerical calculation solely depends on the grid size $\Delta t$, which we 
chose to be  $\omega_0\Delta t=0.01$ in the following. Furthermore we call $t_{\text{max}}=N\Delta t$ the maximal 
evaluated time.

We first consider the dependence of the non-Markovianity on the coupling strength $\gamma$ and on the initial bath 
temperature $T$. In the following plots the initial system states $\rho_1$ and $\rho_2$ are always two differently displaced 
coherent states with variances  $\sigma_{x_{1,2}x_{1,2}} m\omega_0 = \sigma_{p_{1,2}p_{1,2}}/(m\omega_0)=0.5$, 
$\sigma_{p_{1,2}x_{1,2}}=0$.

In Fig.~\ref{longNMgamT_varT} (a) we show the non-Markovianity measure $\mathcal{N}$ in dependence on the 
coupling strength $\gamma$ for different temperatures of the initial bath state and a fixed frequency cutoff 
$\Omega/\omega_0=100$. We observe that $\mathcal{N}$ rapidly decreases as the coupling strength goes to zero.
This can be understood from the fact that in the limiting case $\gamma/\omega_0 \ll 1$ the dynamics is very well
approximated by the quantum optical (weak coupling) master equation for the damped harmonic oscillator which is in 
Lindblad form and, hence, Markovian \cite{Breuer2002}. As can be seen from Fig.~\ref{longNMgamT_varT} (c)
for sufficiently small values of $\gamma/\omega_0$ the non-Markovianity remains small for all temperatures, pointing to
the fact that the Markov approximation is good in this regime uniformly in temperature \cite{Breuer2002}.
At intermediate couplings we see a maximum of the non-Markovianity measure, the height of which decreases for 
higher temperatures. Tuning the coupling to higher values leads to a decrease of $\mathcal{N}$, which is surprising 
on first sight since stronger couplings should increase the ability of information exchange between system and bath. 

\begin{figure}[h]
\centering
\includegraphics[]{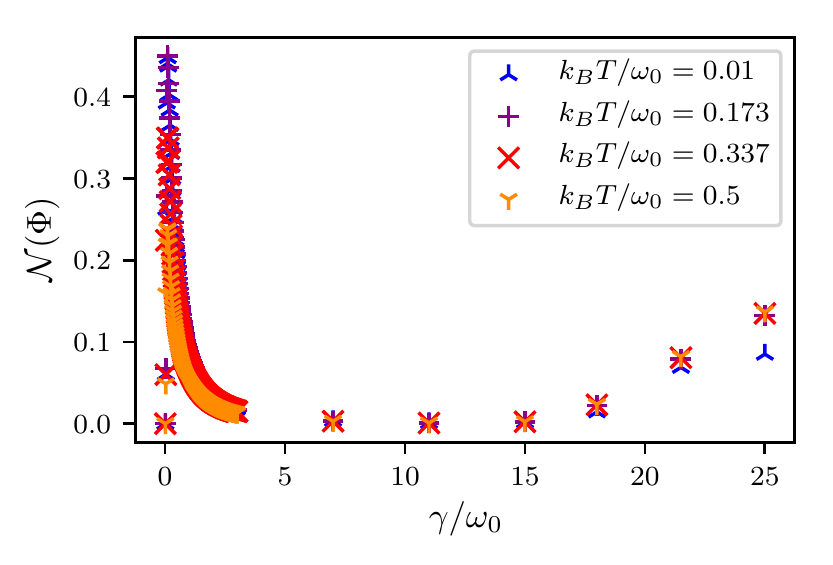}
\caption{The non-Markovianity $\mathcal{N}$ as a function of the coupling strength $\gamma$ for different 
temperatures $T$. The parameters are the same as in Fig.~\ref{longNMgamT_varT} (a), but here we show
a larger range of the coupling to demonstrate the increase of the non-Markovianity for very high damping.}
\label{NMgamhighrange}
\end{figure}

The fact that $\mathcal{N}$ decreases to zero for increasing couplings can be understood from the following
argument which holds in the strong friction Smoluchowski regime given by $\gamma/\omega_0 \gg 1$
and $\gamma/\omega_0 \gg \omega_0/k_BT$ \cite{QuantumSmuAnkerhold,QuantumSmuHanggi}.
First, decoherence times are very small in the strong friction limit which leads to a very rapid decay of the
off-diagonals of the density matrix in the position representation. Moreover, within the Smoluchowski regime 
the diagonals of the density matrix in the position representation are governed by the Smoluchowski
equation which is of the form of a Fokker-Planck equation and, thus, describes a Markovian process.

What happens if one increases the coupling $\gamma$ even further such that one re-enters 
(for a fixed cutoff $\Omega$) the blue region of Fig.~\ref{disc} in which the discriminant is
negative and two complex roots of \eqref{roots} exist? In view of the oscillatory nature of the
solution one should expect that in this regime of ultra strong damping non-Markovian dynamics re-emerges, 
which is indeed confirmed by Fig.~\ref{NMgamhighrange}.

Finally, in Fig.~\ref{longNMgamT_varT}(b) we observe a rapid decrease of $\mathcal{N}$ for high temperatures. 
At very low temperatures our simulations indicate the existence of a maximum of the non-Markovianity as a 
function of temperature. The height of this maximum increases below a certain value of the coupling and decreases again 
for higher couplings.

\begin{figure}
\centering
\includegraphics[]{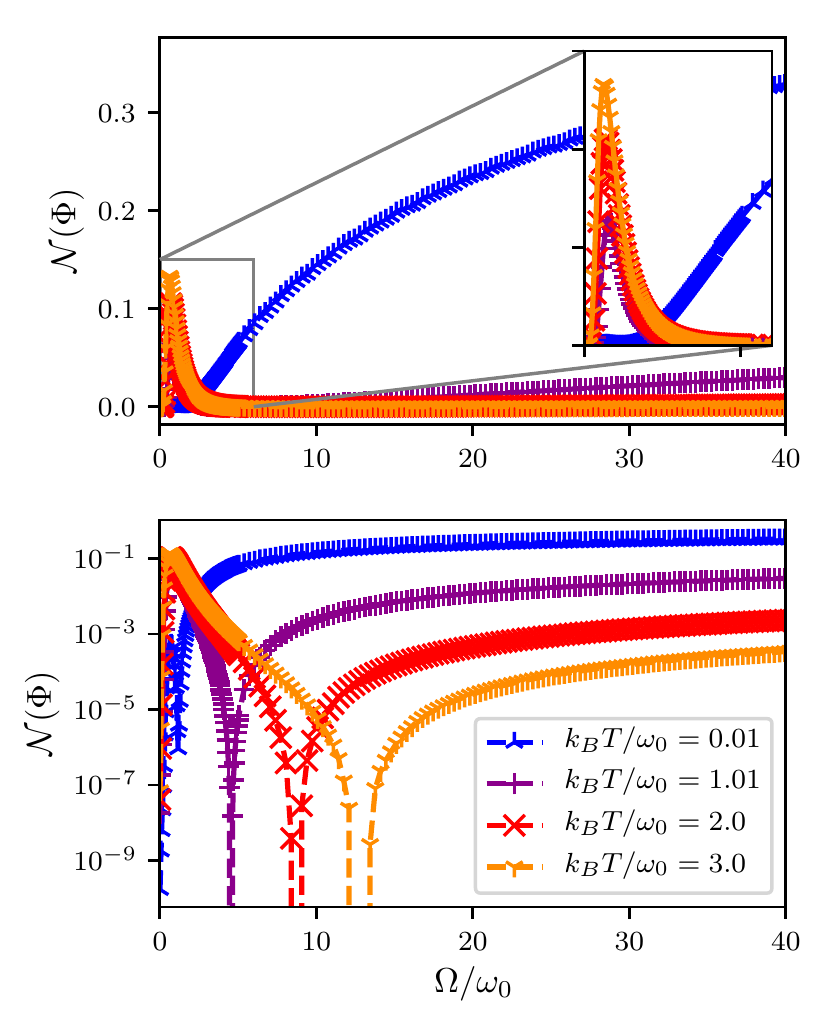}
\caption{Top: Non-Markovianity measure $\mathcal{N}(\Phi)$ in dependence on the cutoff frequency $\Omega$ of the 
Lorentz-Drude spectral density for different initial bath temperatures $T$ and a fixed coupling $\gamma/\omega_0=0.1$. 
The maximal evaluated time is $t_{\text{max}}\omega_0=40$.  The displacements are 
$\langle x_1\rangle\sqrt{2m\omega_0}=-3$, $\langle x_2\rangle\sqrt{2m\omega_0}=3$, as well as 
$\langle p_1\rangle=\langle p_2\rangle=0$. Bottom: The same plot with a logarithmic scale on the vertical axis.} 
\label{longNMomT_varT}
\end{figure}

\begin{figure}
\centering
\includegraphics[]{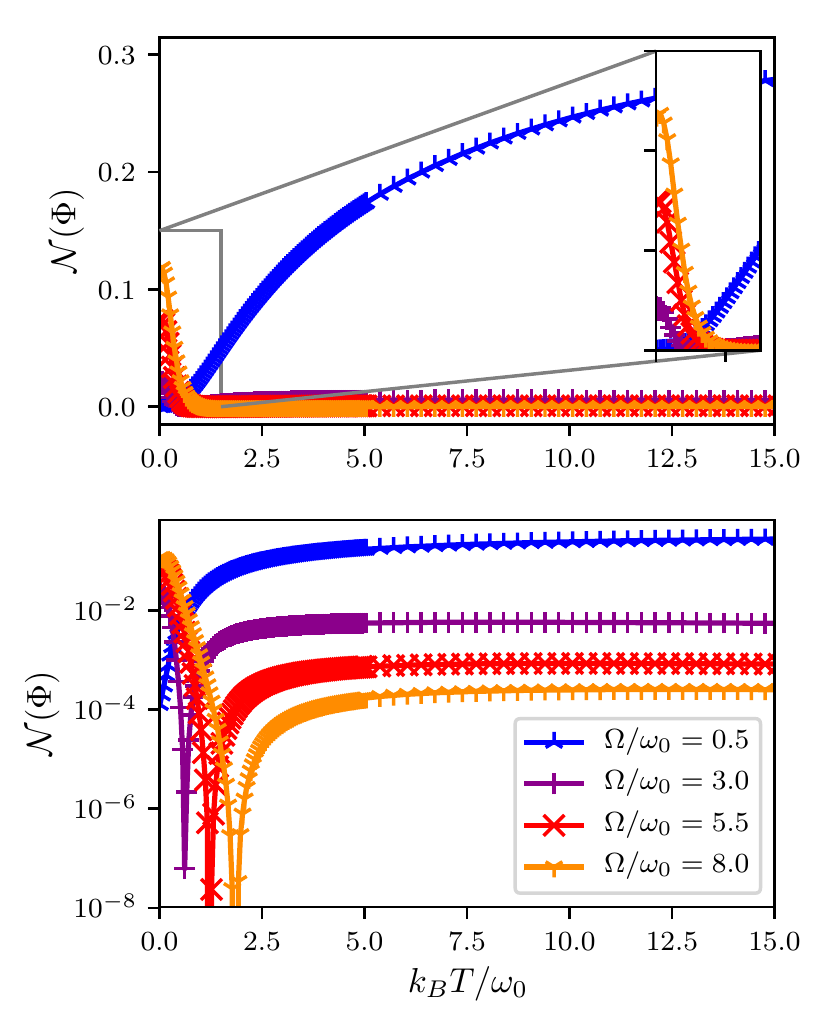}
\caption{Top: Non-Markovianity measure $\mathcal{N}(\Phi)$ in dependence on the initial bath temperatures $T$ for 
different cutoff frequencies $\Omega$ of the Lorentz-Drude spectral density and a fixed coupling $\gamma/\omega_0=0.1$. 
The maximal evaluated time is $t_{\text{max}}\omega_0=40$. The displacements are 
$\langle x_1\rangle\sqrt{2m\omega_0}=-3$, $\langle x_2\rangle\sqrt{2m\omega_0}=3$, as well as 
$\langle p_1\rangle=\langle p_2\rangle=0$.
Bottom: The same plot with a logarithmic scale on the vertical axis.}
\label{longNMomT_varom}
\end{figure}

Next, we consider the non-Markovianity in dependence on the cutoff frequency $\Omega$ and the initial 
bath temperature $T$.
Figure \ref{longNMomT_varT} shows the non-Markovianity as a function of the cutoff frequency $\Omega$
for different initial bath temperatures $T$. The system-bath coupling is fixed to $\gamma/\omega_0=0.1$. 
The first point one notices in the upper plot of Fig.~\ref{longNMomT_varT} is that there exists a 
local maximum not far from the cutoff frequency $\Omega/\omega_0=1$, which increases and slightly shifts to lower cutoffs 
for higher temperatures. In the bottom plot we show the same graphs on a 
logarithmic scale in the vertical axis. This also reveals a minimum of the non-Markovianity measure, which moves 
to higher cutoffs and gets broader for increasing temperatures. We will come back to an explanation of this
feature below.

\begin{figure}
\begin{minipage}{0.9\linewidth}
\includegraphics[width=\linewidth]{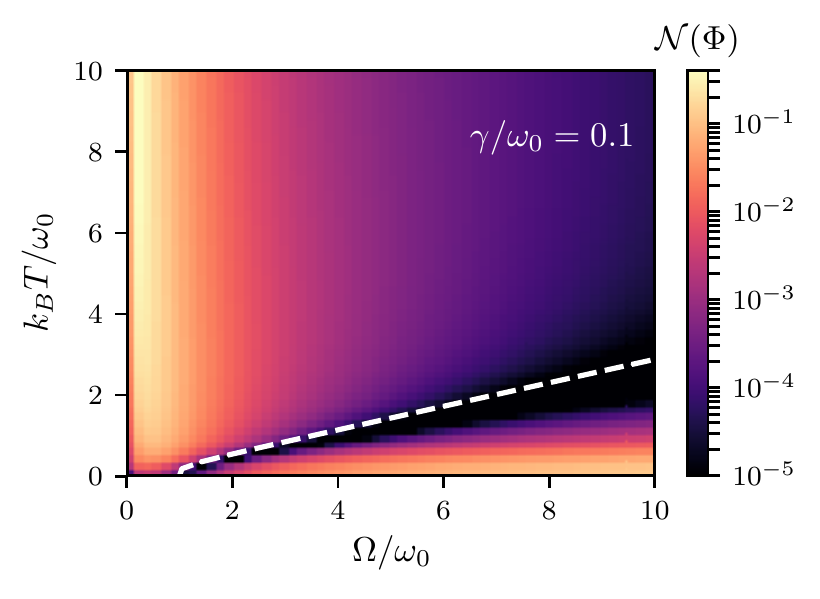}
\end{minipage}
\begin{minipage}{0.9\linewidth}
\includegraphics[width=\linewidth]{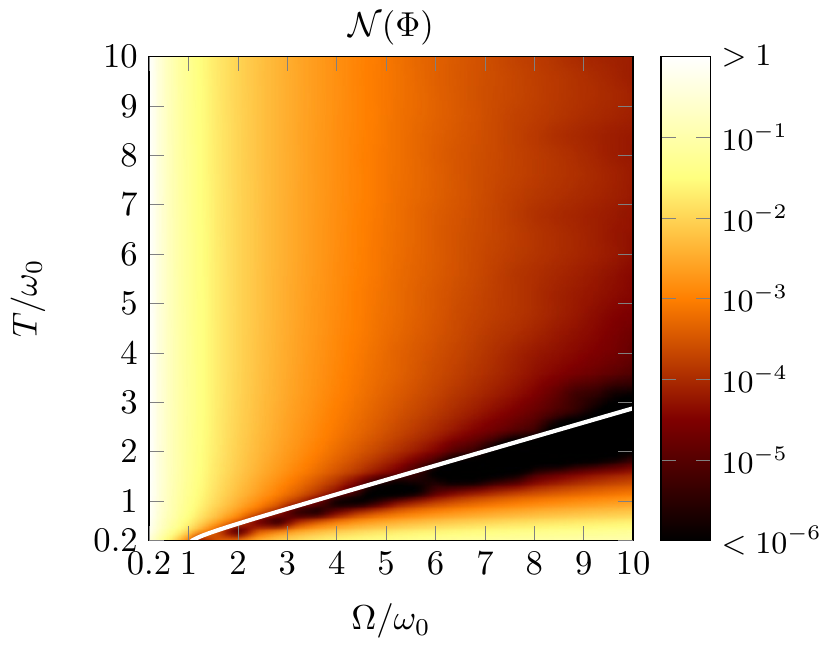}
\end{minipage}
\caption{Top: Non-Markovianity measure for the Caldeira-Leggett model in dependence on $\gamma$ and $\Omega$. 
The initial states were two coherent states with displacements $\langle x_1\rangle\sqrt{2m\omega_0}=-3$, 
$\langle x_2\rangle\sqrt{2m\omega_0}=3$, and $\langle p_1\rangle=\langle p_2\rangle=0$. The maximal evaluated 
time is $t_{\text{max}}\omega_0=25$. The coupling is fixed to $\gamma/\omega_0=0.1$
Bottom: Non-Markovianity measure defined with help of the trace distance in dependence on temperature $T$ and
cutoff frequency $\Omega$ for the Spin-Boson model by the second-order Born approximation \cite{Clos2012a}. The 
coupling was fixed to $\gamma/\omega_0=0.1$. The initial states of the spin correspond to antipodal points of the Bloch 
sphere. In both plots the white curve represents the resonance condition of Eq.~\eqref{resonance-2}. 
\label{fig:tedopaFig}}
\label{2dplots}
\end{figure}

In Fig.~\ref{longNMomT_varom} we present  the non-Markovianity measure $\mathcal{N}$ in dependence on the initial 
bath temperature $T$ for different cutoff frequencies $\Omega$ and a fixed coupling  $\gamma/\omega_0=0.1$. 
Also here we can see from the upper plot a maximum at very low temperatures. After this maximum
$\mathcal{N}$ drops rapidly for increasing temperature. The bottom plot, where the vertical axis has a logarithmic scale, 
reveals again a minimum of $\mathcal{N}$ which moves to higher temperatures for increasing cutoffs.
The dependence of this minimum on the cutoff and the temperature can be seen in the top plot of Fig.~\ref{2dplots}. In the 
bottom plot of Fig.~\ref{2dplots} we show for a comparison the results of a similar study on the non-Markovianity measure in 
the spin-boson model \cite{Clos2012a}. The authors of this study investigated the originally proposed non-Markovianity 
measure involving the trace distance and carried out the maximization procedure by a random sampling of the initial states 
over the whole Bloch sphere. Furthermore, they utilized the second-order Born approximation of the master equation 
(without the Markov approximation) in order determine the dynamics of the spin. It is quite remarkable that,
despite all these differences $\mathcal{N}(\Omega,T)$ behaves almost the same in both models. The main reasons
for these striking similarities are on the one hand the close relationship between the trace distance and the Bures 
distance and, on the other hand, the fact that both models describe an open system with a single transition frequency
$\omega_0$, a linear system-bath interaction and the same spectral density.

In order to explain the cutoff and temperature dependent minimum of the non-Markovianity discussed above
we employ an argument developed in Ref.~\cite{Clos2012a}. This argument is based on the effective spectral
density
\begin{equation}
\label{Jeffective}
J_{\text{eff}}(\omega,\Omega,T)=J(\omega,\Omega)\coth\left(\frac{\omega}{2k_BT} \right)
\end{equation}  
which determines the frequency spectrum of the noise kernel defined in Eq.~(\ref{correlationfunction}). If the effective
spectral density is maximal at $\omega=\omega_0$ the oscillator `feels' a spectral density which is approximately
constant in the vicinity of its eigenfrequency $\omega_0$ and, therefore, approximately corresponds to the spectrum of
white noise favoring Markovian behavior. The idea is thus that
\begin{equation} \label{resonance-1}
 \left. \frac{\partial J_{\text{eff}}(\omega,\Omega,T)}{\partial \omega} \right|_{\omega=\omega_0} = 0
\end{equation}  
is an appropriate condition for predominantly Markovian dynamics. It is easy to verify that this resonance
condition yields a curve in the $(\Omega,T)$ plane which is given by
\begin{equation} \label{resonance-2}
 \Omega = \omega_0
\sqrt{\frac{k_BT \sinh\left(\frac{\omega_0}{k_BT}\right)+\omega_0}{k_BT\sinh\left(\frac{\omega_0}{k_BT}\right)-\omega_0}}
\end{equation}
and shown as white curves in Fig.~\ref{2dplots}. As can be seen from the figure the resonance condition
very well reproduces the location of the minima of the non-Markovinaity measure.

\subsection{Influence of external driving on non-Markovianity}\label{sec:driving}

Finally, we investigate the impact of an external force on the non-Markovianity measure, driving the open system 
and/or the bath modes. Remarkably, it turns out that driving does not influence the non-Markovianity in our case. 
This is in contrast to other studies, which have investigated the influence of driving on a similar measure of 
non-Markovianity on an open two-level systems, for example in Ref. \cite{NMdrivingPoggi}. In \cite{NMdrivingSampaio} it 
was shown that the system driving can even increase non-Markovian effects.

A general driving of the system or bath modes results in an additional term in the Hamiltonian (\ref{ham}) of the form
\begin{equation}
\label{Hext}
H_{\text{ext}}(t)=-\left(d_0 x+\sum_{n} d_n x_n\right)F(t), 
\end{equation}
where the $d_{\alpha}$ represent the driving strengths of the system and the bath modes and $F(t)$ is the external force.
It is straightforward to repeat the derivation of the Heisenberg equation (\ref{quantlangevin}) 
taking into account the terms stemming from the external Hamiltonian contribution (\ref{Hext}). The result is
\begin{equation}
\ddot{x}(t)+\omega_0^2 x(t)+\frac{d}{dt}\int\limits_0^t dt^{\prime} \gamma(t-t^{\prime})x(t^{\prime})=\frac{1}{m}(B(t)+F_{\text{eff}}(t)),
\end{equation}
where
\begin{equation}
F_{\text{eff}}(t)=d_0 F(t)+\int\limits_0^t dt^{\prime} \Lambda(t-t^{\prime}) F(t^{\prime})
\end{equation}
with $\Lambda(t)=\sum_{n} \frac{d_n \kappa_n}{m_n \omega_n} \sin\left(\omega_n t\right)$
is the effective driving force consisting of the direct external driving of the system and the indirect driving 
mediated by the bath \cite{DrivenCLGrabert}.
We again solve this equation via a Laplace transformation. This results in an additional term
\begin{equation}
 x_D(t)=\frac{1}{m}\int\limits_0^t dt^{\prime} G_2(t-t^{\prime}) F_{\text{eff}}(t^{\prime})
\end{equation}
 to the previously obtained solution of Eq. (\ref{generalsol}), which we now call $x_H(t)$ such that
 \begin{equation}
 x(t)=x_H(t)+x_D(t).
\end{equation}
Since the driving term $x_D(t)$ is a c-number function it follows that
\begin{align}
\big\langle x_D^2(t) \big\rangle &=\big\langle x_D(t) \big\rangle^2,\\
\big\langle x_H(t)x_D(t)\big\rangle &=\big\langle x_H(t)\big\rangle \big\langle x_D(t)\big\rangle,
\end{align}  
which means that the driving term has no influence on the position variance. 
Analogously, one can see that the driving term has also no influence on the other elements of the covariance matrix.
Still, the mean values of the position and momentum are affected and are given by
\begin{align}
\big\langle x(t) \big\rangle &=\big\langle x_H(t) \big\rangle +\int\limits_0^t dt^{\prime} G_2(t-t^{\prime}) 
F_{\text{eff}}(t^{\prime}),\\
\big\langle p(t) \big\rangle &=\big\langle p_H(t) \big\rangle +\int\limits_0^t dt^{\prime} \dot{G}_2(t-
t^{\prime}) F_{\text{eff}}(t^{\prime}).
\end{align}
However, in the expression (\ref{Fidelitygauss}) for the quantum fidelity of Gaussian states only the difference between the 
mean values $\mathbf{d}_2-\mathbf{d}_1$ enters, such that the driving contributions cancel each other since they are 
independent of the initial conditions. Hence, the fidelity is not influenced by the driving of the system and/or bath modes 
and thus also the non-Markovianity measure remains unaffected. The above derivation shows that the main
reasons for this fact are the Gaussian nature of the initial states and the linearity of the model.

\section{Conclusions}\label{sec:conclu}

In this article we investigated the emergence of memory effects in the Caldeira-Leggett model for different 
parameter regimes of the coupling strength, the frequency cutoff of the spectral density and the bath temperature. 
We quantified non-Markovianity with help of a measure especially suitable for continuous variable systems, 
which is based on the Bures metric as distance measure for quantum states. We used Gaussian initial states for the open 
system which allowed us to exactly determine the time evolution for arbitrary parameter combinations without any 
approximation.

Our main results may be summarized as follows.

\begin{itemize}

\item[1.]
We found that the non-Markovianity measure exhibits a non-monotonic behaviour as a function of the
coupling strength and the bath temperature. In particular, it shows a maximum at intermediate coupling strengths 
and decreases for higher values of the latter (see Sec.~\ref{sec:non-Markoviantiy}).

\item[2.]
For small coupling strength the non-Markovianity measure as a function of frequency cutoff and temperature shows 
remarkable similarities to the spin-boson model for which an analogous study is discussed in Ref.~\cite{Clos2012a}. 
In both models one finds a pronounced Markovian area (minimum of the non-Markovianity measure) inside a 
non-Markovian regime which can be understood as a resonance effect resulting from the maximum of an effective,
temperature-dependent spectral density (see, in particular, Eq.~\eqref{resonance-1}).

\item[3.] The decrease of the non-Markovianity with increasing coupling strength in the regime of strong dissipation
can be explained as resulting from the dynamical behavior in the Smoluchowski regime
\cite{QuantumSmuAnkerhold,QuantumSmuHanggi}. In addition, we observe
an anew increase of the non-Markovianity for ultra strong coupling (see Sec.~\ref{sec:non-Markoviantiy}).

\item[4.] We showed that non-Markovianity cannot be generated within the approximation provided
by the Caldeira-Leggett master equation for Gaussian initial states (see Sec.~\ref{sec:bures}).

\item[5.] Finally, we demonstrated that a driving of the system and/or the bath modes by an arbitrary 
external force has no influence on the non-Markovianity of the open system dynamics which is due to
the linearity of the model and the use of Gaussian initial states (see Sec.~\ref{sec:driving}).

\end{itemize}

In summary, we observed a rich behavior of the non-Markovianity of quantum Brownian motion, even though
our model is linear and thus integrable. Several of the features described obviously depend on the linearity and
on the Gaussian character of the initial states. We expect that further interesting phenomena emerge if one
considers a nonlinear open system or non-Gaussian initial states. Another interesting problem is to study
further the relations between different physical models with regard to their degree of non-Markovianity. Our
findings on the similarity between quantum Brownian motion and the spin-boson model point to a kind of
universality of non-Markovian behavior which should be scrutinized in more detail.

\begin{acknowledgments}
AK  acknowledges support from the Georg H. Endress foundation.
We further acknowledge support from the state of Baden-W\"urttemberg by bwHPC.
\end{acknowledgments}

\appendix

\section{Calculation of the Noise contributions}\label{appA}

We illustrate the procedure of determining the noise integrals \eqref{noise1}-\eqref{noise3} by means of 
$\langle I_{xx}(t)\rangle$ in the following.
We have
\begin{equation}
\langle I_{xx}(t)\rangle=\frac{2\gamma k_B T \Omega^2}{m}\sum_{n=-\infty}^{\infty}\frac{1}{\Omega^2-\nu_n^2} 
\mathcal{I}(t,n),
\end{equation} where
\begin{widetext}
\begin{align}
\mathcal{I}(t,n)&= \int^t_0 ds\int^t_0 ds^{\prime}G_2(t-s)G_2(t-s^{\prime})\left(\Omega e^{-\Omega |s-s^{\prime}|}-|\nu_n| 
e^{-|\nu_n||s-s^{\prime}|}\right)\\
&=\sum_{i,j=1}^{3}d_i d_j e^{(z_i+z_j)t}\int^t_0 ds\int^t_0 ds^{\prime} e^{-(z_i s+z_j s^{\prime})}\left(\Omega e^{-\Omega |s-
s^{\prime}|}-|\nu_n| e^{-|\nu_n||s-s^{\prime}|}\right). 
\end{align}
We change to the new integration variables $\tilde{s}=s-s^{\prime}$ and $\tau=s^{\prime}$. This leads to 
\begin{align}
\mathcal{I}(t,n)=\sum_{i,j=1}^{3}d_i d_j e^{(z_i+z_j)t}\int^t_0 d\tau\int^{t-\tau}_{-\tau} d\tilde{s} e^{-(z_i (\tilde{s}+\tau)+z_j 
\tau)}\left(\Omega e^{-\Omega |\tilde{s}|}-|\nu_n| e^{-|\nu_n||\tilde{s}|}\right).
\end{align}
The time integrals can be evaluated straightforwardly to obtain the expression
\begin{equation}
\begin{split}
\mathcal{I}(t,n)=\sum_{i,j=1}^{3}d_i d_j &\Bigg\{\frac{|\nu_n|}{|\nu_n|-z_i}\bigg(\frac{1-e^{(z_i+z_j)t}}{z_i+z_j}-\frac{e^{(-|
\nu_n|+z_i)t}-e^{(z_i+z_j)t}}{|\nu_n|+z_j}\bigg) \\
&-\frac{\Omega}{\Omega-z_i}\bigg(\frac{1-e^{(z_i+z_j)t}}{z_i+z_j}-\frac{e^{(-\Omega+z_i)t}-e^{(z_i+z_j)t}}{\Omega+z_j}\bigg)
\\
&+\frac{|\nu_n|}{|\nu_n|+z_i}\bigg(\frac{1-e^{(z_j-|\nu_n|)t}}{|\nu_n|-z_j}+\frac{1-e^{(z_i+z_j)t}}{z_i+z_j}\bigg) 
-\frac{\Omega}{\Omega+z_i}\bigg(\frac{1-e^{(z_j-\Omega)t}}{\Omega-z_j}+\frac{1-e^{(z_i+z_j)t}}{z_i+z_j}\bigg)\Bigg\}.
\end{split}
\end{equation}
\end{widetext}

\bibliography{biblio_nonmarkov}

\end{document}